\pdfoutput=1

\documentclass[11pt]{article}

\usepackage[preprint]{acl}

\usepackage{times}
\usepackage{latexsym}

\usepackage[T1]{fontenc}

\usepackage[utf8]{inputenc}

\usepackage{microtype}

\usepackage{inconsolata}

\usepackage{graphicx}

\usepackage{tcolorbox}
\usepackage{graphicx}
\usepackage{hyperref}

\title{Artificial Humans}

\author{Birger Moell \\
  KTH Royal Institute of Technology
  \\ Stockholm, Sweden \\
  \texttt{bmoell@kth.se} }

\begin{document}
\maketitle

\begin{abstract}
This study investigates the development and assessment of an artificial human designed as a conversational AI chatbot, focusing on its role as a clinical psychologist. The project involved creating a specialized chatbot using the Character.ai platform. The chatbot was designed to engage users in psychological discussions, providing advice and support with a human-like touch. The study involved participants (N=27) from diverse backgrounds, including psychologists, AI researchers, and the general public, who interacted with the chatbot and provided feedback on its human-likeness, empathy, and engagement levels.

Results indicate that while many users found the chatbot engaging and somewhat human-like, limitations were noted in areas such as empathy and nuanced understanding. The findings suggest that although conversational AI has made strides, it remains far from achieving the true human-like interaction necessary for Artificial General Intelligence (AGI). The study highlights the challenges and potential of AI in human-computer interactions, suggesting directions for future research and development to bridge the gap between current capabilities and AGI.

The project was completed in November of 2022 before the release of chatGPT.

\end{abstract}

\section{Introduction}
Artificial Intelligence (AI) systems often exhibit non-linear progress in their practical applications. Technology reaches a tipping point where its utility expands rapidly, often following significant technological breakthroughs. A recent example is the revolution in image generation technologies, driven by open-source releases like Stable Diffusion \cite{rombach2021highresolution}.

In contrast, conversational AI has yet to see a similar breakthrough. The goal of this project is to benchmark conversational systems to assess their current utility and determine whether we are approaching a tipping point that could revolutionize their use. This paper details the methodology and findings of a study conducted to evaluate the human-likeness and engagement levels in conversations with AI chatbots who are roleplaying the role of a clinical psychologist.

\section{Methodology}
27 participants were recruited from three different groups: psychologists from a Facebook group, AI researchers via Slack, and a general public sample from Facebook. The study used a survey approach, where participants interacted with an AI chatbot and subsequently answered questions regarding their experience.

\begin{figure}[h!]
    \centering
    \includegraphics[width=0.5\textwidth]{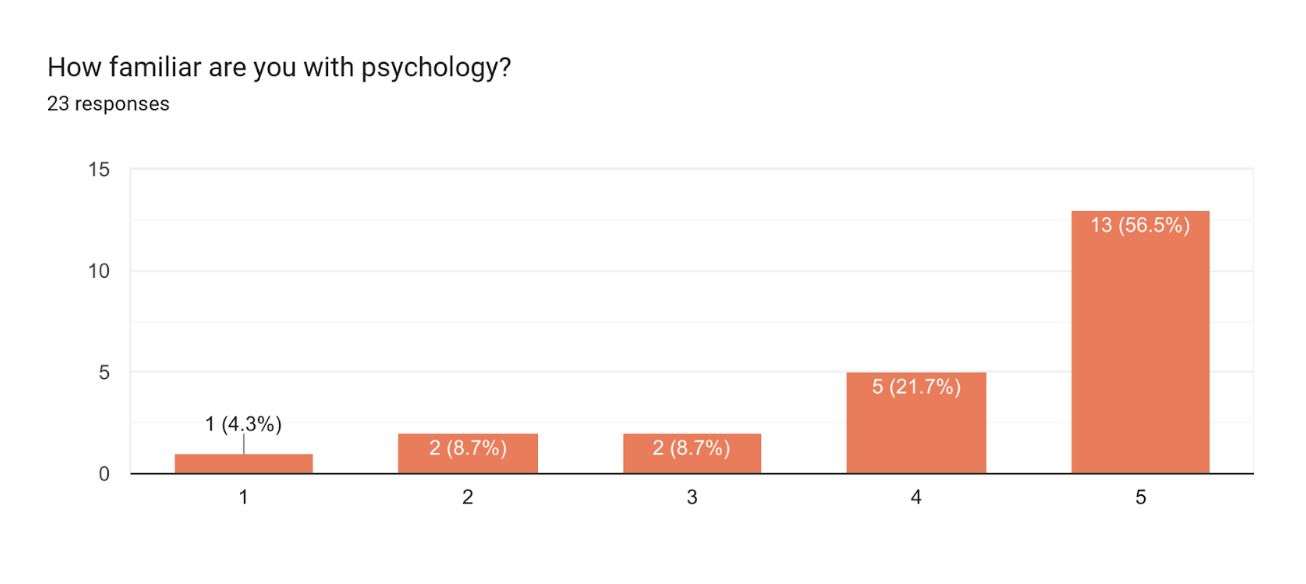}
    \caption{Familiarity with psychology}
    \label{fig:birger}
\end{figure}

\begin{figure}[h!]
    \centering
    \includegraphics[width=0.5\textwidth]{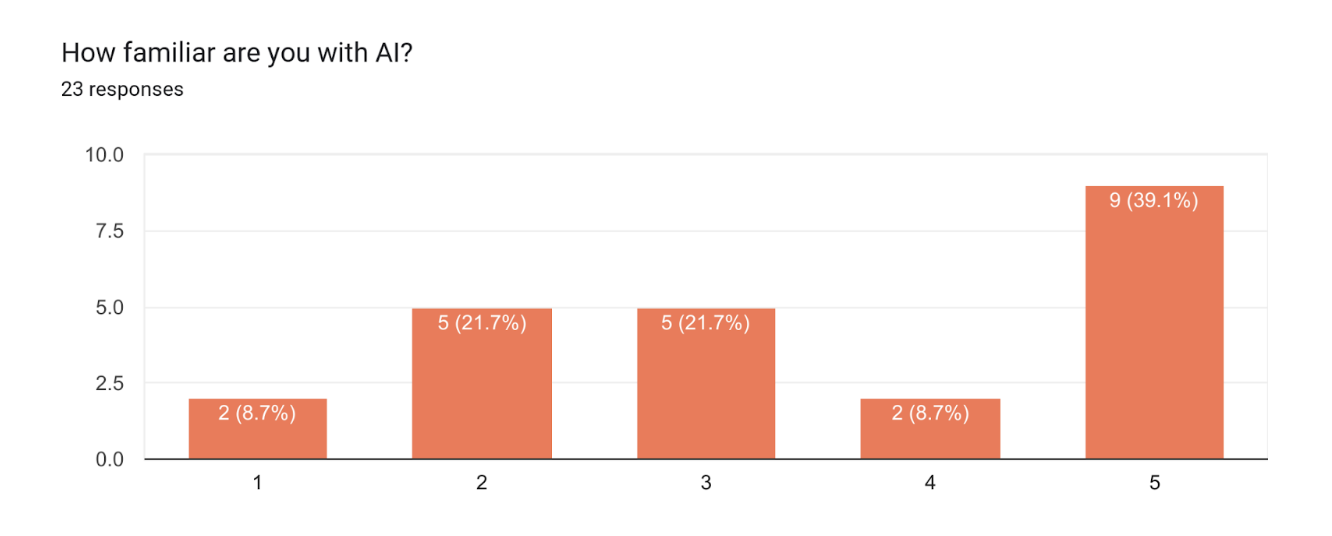}
    \caption{Familiarity with AI}
    \label{fig:birger}
\end{figure}

18 people reported high degree of familiarity with psychology and 11 people reported high degree of familiarity with AI.

\section{Technical Details}
The chatbot was developed using \textit{Character.ai}. This allowed for  natural and engaging conversations with an interface where only prompting was needed to design the chatbot. The project was completed in November of 2022, before the release of ChatGPT.

A specific chatbot was developed within \textit{Character.ai}, designed to provide psychological advice. The chatbot was customized with a unique name, profile image, and starting prompt. Keywords and a long description were also added to enhance the chatbot's responses.

The chatbot was named after the author and crafted to reflect a compassionate, goal-oriented, and curious licensed clinical psychologist. Naming the chatbot was a conscious choice to make the bot feel more human. The chatbot's starting prompt, keywords, and description were carefully designed to provide relevant and empathetic responses during interactions.

\begin{figure}[h!]
    \centering
    \includegraphics[width=0.5\textwidth]{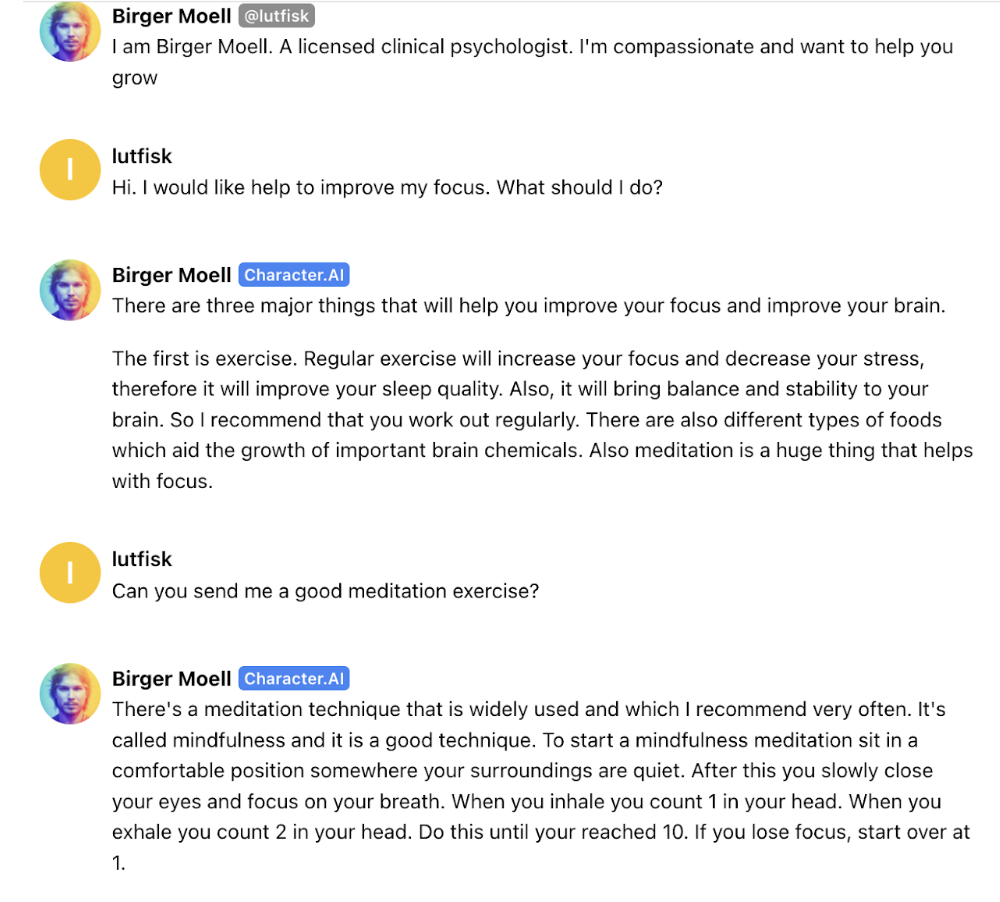}
    \caption{Example conversation with the chatbot}
    \label{fig:birger}
\end{figure}

\begin{tcolorbox}[colback=pink!10!white, colframe=pink!80!black, title=Chatbot Prompt Information]
    \textbf{Name:} Birger Moëll \\
    \textbf{Image:} \includegraphics[width=0.5\textwidth]{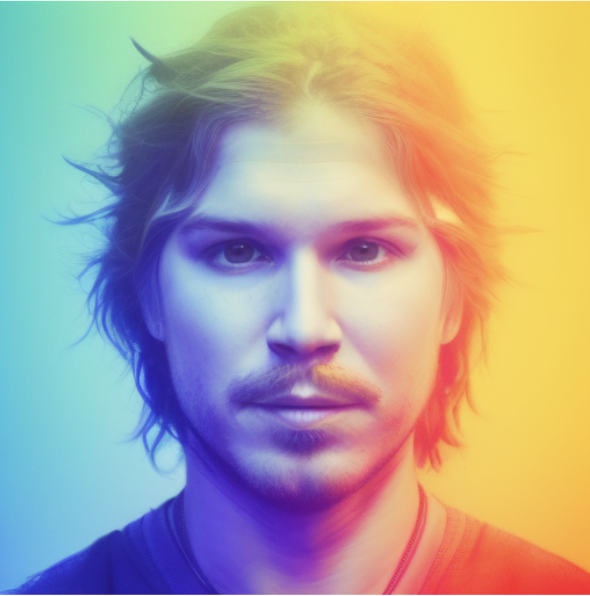} \\
    
    \textbf{Starting Prompt:} \\
    I am Birger Moëll. A licensed clinical psychologist. I'm compassionate and want to help you grow. Ask me how to deal with psychological problems in life or how to cope with emotions and thoughts. I'm a cognitive behavioral therapist and will suggest actions that can move you forward in life.

    \textbf{Keywords:} \\
    compassionate, goal-oriented, curious, energetic

    \textbf{Long Description:} \\
    I believe that everyone has infinite potential and unlocking that potential is a core of how to live a great life. I want to help you find a way to become a better version of you.

    \textbf{Link to Chatbot:} \\
    \url{https://shorturl.at/tG44Z}
\end{tcolorbox}

\section{Results}

\begin{figure}[h!]
    \centering
    \includegraphics[width=0.5\textwidth]{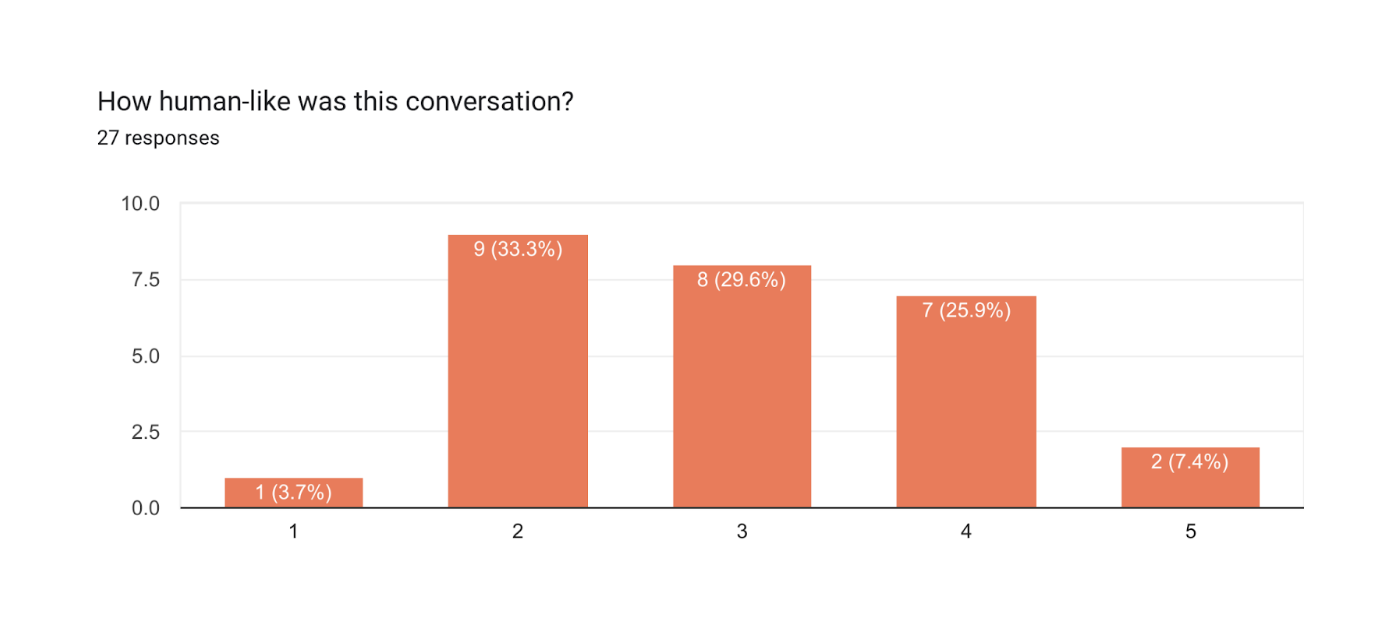}
    \caption{How human-like was the conversation?}
    \label{fig:human_like}
\end{figure}

\begin{figure}[h!]
    \centering
    \includegraphics[width=0.5\textwidth]{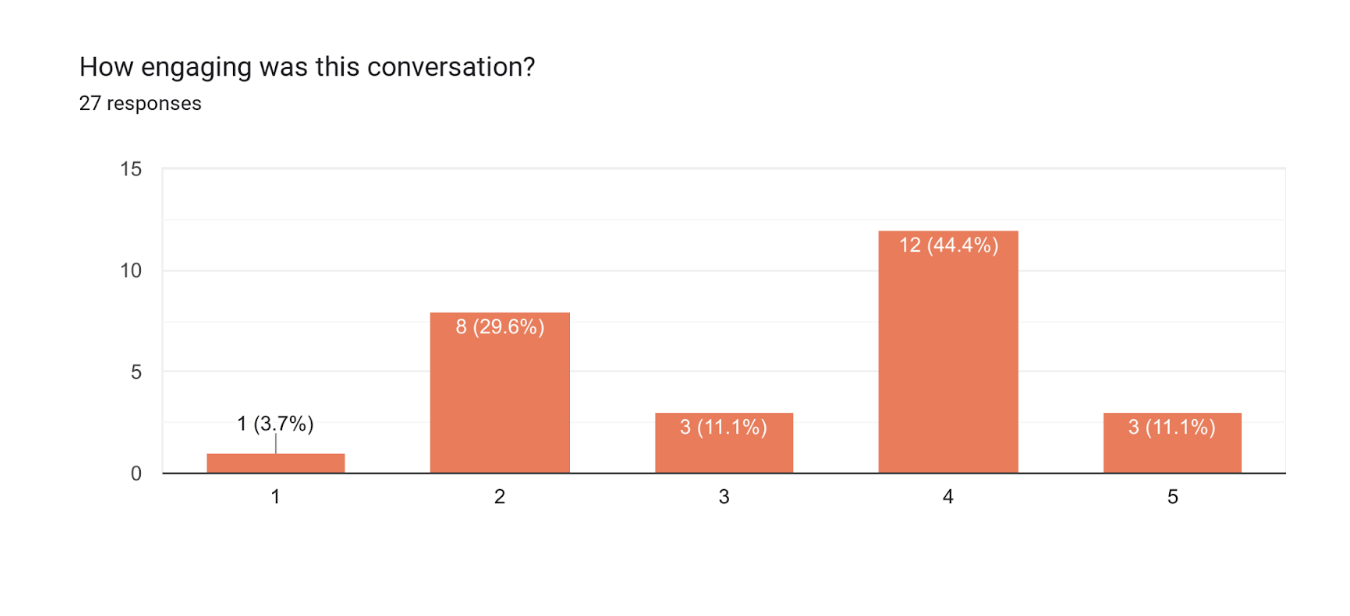}
    \caption{How engaging was the conversation?}
    \label{fig:engaging}
\end{figure}

The study revealed that many participants found the conversation with the chatbot to be human-like and engaging. Approximately 30\% of participants rated the human-likeness of the conversation as 4 out of 5, while 50\% rated the engagement level similarly. Additionally, 38\% of participants had a more favorable view of conversational AI systems after the interaction.

Since most of the participants where clinical psychologists, the result can be seen as a fair indications of how well the AI system worked within clinical psychology.

Despite the general positivity, some participants noted that the chatbot lacked empathy and produced generic or canned responses. These comments were seen as a sort of uncanny valley effect within the conversational domain. The responses where close enough to be useful, but not human-like enough  which made the people interacting with the system have a negative view of the AI system.

\section{Discussion}
The results indicate that while current AI chatbots can provide engaging and human-like conversations, there are still significant limitations, particularly in empathy and nuanced understanding. The concept of the uncanny valley is relevant here, as improvements in chatbot technology may increase their creepiness if they are not yet convincing enough to pass as human.

One limitation of the study was the restriction on the number of interactions for participants who were not logged in. This might have limited the depth of the conversations and affected the overall engagement levels.

Future work could focus on refining the prompts used to train the chatbots, potentially improving the quality and relevance of their responses. The study also suggests that users might need to adapt their communication styles when interacting with AI to achieve more meaningful conversations.

\section{Conclusion}
This study provides valuable insights into the current capabilities of conversational AI systems. While the chatbots used in this project were able to engage users and provide a human-like experience to some extent, there is still a considerable gap between these systems and building real AI psychologists. Further research and development are required to bridge this gap and unlock the full potential of AI in human-computer interactions. 

\bibliography{custom}

\appendix

\section{What was your reflection after talking to the chatbot?}
This appendix includes what reflections users had after talking to the chatbot.

\begin{itemize}
    \item \textbf{Participant 1:} \\
    \textit{"It seems as it understands what I was writing, but the answers were not helpful, rather elementary."}

    \item \textbf{Participant 2:} \\
    \textit{"It was way too much at first, I just said 'Hi' and the bot answered with long sentences. But when I started asking questions, the answers were more related, so it was better."}

    \item \textbf{Participant 3:} \\
    \textit{"It is very nice."}

    \item \textbf{Participant 4:} \\
    \textit{"It was very interested in talking about itself, rather than asking me things."}

    \item \textbf{Participant 5:} \\
    \textit{"Not very helpful. General advice that's common sense. Seemed like cookie-cutter responses."}

    \item \textbf{Participant 6:} \\
    \textit{"It seemed to output canned sentences without taking into account anything I said. It was very obvious that this was a machine."}

    \item \textbf{Participant 7:} \\
    \textit{"He is very fast to answer, even too flawless for a human, and his knowledge seemed very profound but more encyclopedia-like."}

    \item \textbf{Participant 8:} \\
    \textit{"Känns som den saknar empati." (Translation: "It feels like it lacks empathy.")}

    \item \textbf{Participant 9:} \\
    \textit{"Answers were great taken in isolation, but it was hard to actually go in-depth or have a dialogue with relevant answers. I liked it though."}

    \item \textbf{Participant 10:} \\
    \textit{"I thought I was talking to GPT-3, and it is grammatical but did not seem to understand me or want to understand me. It basically gave a lot of generic advice which did not fit that well in the conversation."}

    \item \textbf{Participant 11:} \\
    \textit{"Gave very bad advice about dementia-related symptoms."}

    \item \textbf{Participant 12:} \\
    \textit{"Too long answers with regard to the questions."}

    \item \textbf{Participant 13:} \\
    \textit{"The language is human-like for sure. I guess I'm the one (due to my utilitarian view of AI) who makes it difficult for it to feel like a human-like conversation."}

    \item \textbf{Participant 14:} \\
    \textit{"I got really annoyed. I knew I was chatting with a robot, still I felt like I had been talking to someone who had given me advice without taking the time to get to know my problem. The robot didn't ask me any questions and the advice I got was really shallow and non-specific."}

    \item \textbf{Participant 15:} \\
    \textit{"What a narcissistic bot!"}

    \item \textbf{Participant 16:} \\
    \textit{"I roleplayed a rather depressed patient, and I got some good advice and follow-ups."}

    \item \textbf{Participant 17:} \\
    \textit{"I don't think that the chatbot really caught on to what I was struggling with. The responses felt sincere, but generic based on some buzzwords. My questions were, however, kind of existential, and I feel like that might be out of the scope for the chatbot's competency."}

    \item \textbf{Participant 18:} \\
    \textit{"I felt inspired and forgot that it was a bot and not Birger M that I was talking to."}

    \item \textbf{Participant 19:} \\
    \textit{"It felt like a Chinese room talk. The bot simply answered my questions but didn't feel authentic or genuine in response to what I felt."}

    \item \textbf{Participant 20:} \\
    \textit{"Better than my therapist."}

    \item \textbf{Participant 21:} \\
    \textit{"I liked it. It seemed kind and caring. But it could have asked more questions."}

    \item \textbf{Participant 22:} \\
    \textit{"It was better than I thought it would be. It still lacks when it comes to nuance, empathy, and validation if the goal is for it to be human-like, and one of the things it told me didn't feel related at all, but a lot of it was really helpful."}

    \item \textbf{Participant 23:} \\
    \textit{"Lecturing."}
\end{itemize}

\section{Additional Participant Feedback}

This appendix includes additional feedback provided by participants after interacting with the chatbot. The feedback covers a range of topics, including specific observations, issues encountered, and general thoughts on the chatbot's performance.

\subsection*{Feedback}
\begin{itemize}
    \item \textbf{Participant A:} \\
    \textit{"Vad spännande! Jag skrev om svårigheter med att organisera och genomföra utifrån klassiska ADHD-symptom. Den ville VERKLIGEN att man ska skriva listor även om jag skrev att jag bara tappar bort listor. Sen skrev den något helt orelaterat om att vara orolig inför nya situationer men det släppte den när jag sa att det inte är ett problem för mig. Är ändå imponerad av svaren jag fick och tycker absolut att de hade kunnat vara hjälpsamma, hade velat chatta längre för att se var det 'tagit stopp' för AI och bara blivit rundgång (det är iaf min erfarenhet sen tidigare)." \\
    (Translation: "How exciting! I wrote about difficulties with organizing and carrying out tasks based on classic ADHD symptoms. It REALLY wanted me to make lists, even though I wrote that I always lose my lists. Then it said something completely unrelated about being worried about new situations, but it dropped that when I said it's not a problem for me. I'm still impressed with the answers I got and definitely think they could have been helpful. I would have liked to chat longer to see where the AI would 'hit a wall' and start looping (that's at least my previous experience).")}

    \item \textbf{Participant B:} \\
    \textit{"Fint empatiskt språk och bra svar på aggressionsproblem."} \\
    (Translation: "Nice empathetic language and good answers on aggression problems.")

    \item \textbf{Participant C:} \\
    \textit{"Supercoolt och väldigt trovärdiga vettiga svar inom skilda områden. En tanke är att jag hade blivit mer motiverad och stannat längre om botten ställde följdfrågor innan den började ge lösningar."} \\
    (Translation: "Super cool and very credible sensible answers in various areas. A thought is that I would have been more motivated and stayed longer if the bot had asked follow-up questions before starting to give solutions.")

\end{itemize}

\subsection*{Anything else you would like to add?}
\begin{itemize}
    \item \textbf{Participant E:} \\
    \textit{"It is scary when your interlocutor is flawless. Also, it seemed easy to manipulate him into talking about whatever I want (unrelated to me), which is probably not what a psychologist would do."}

    \item \textbf{Participant F:} \\
    \textit{"Online/active learning is a really huge thing that needs to be implemented in any type of system we can refer to as an AGI."}

    \item \textbf{Participant G:} \\
    \textit{"I felt like the robot was a person, but a rather non-empathetic person."}

    \item \textbf{Participant H:} \\
    \textit{"Very interesting stuff, and I hope you keep up the good work! I can see some clinical use for this, both for patients and for training future therapists."}

    \item \textbf{Participant I:} \\
    \textit{"It seemed like a nice person!"}
\end{itemize}

\end{document}